\documentclass[12pt]{iopart}
\usepackage{iopams}  
\usepackage{graphicx}
\usepackage{dcolumn}
\usepackage{bm}
\usepackage{hyperref}
\usepackage{cite}

\usepackage{rotating}
\usepackage{array}
\usepackage{amssymb}
\usepackage{tikz}

\def\be{\begin{equation}}
\def\ee{\end{equation}}
\def\bea{\begin{eqnarray}}
\def\eea{\end{eqnarray}}
\def\ahalf{{\textstyle{1\over2}}}

\def\ncal{\mbox{$\cal N\,$}}

\def\<{\langle}
\def\>{\rangle}

\begin{document}
\nocite{*}

\title{Fractal position spectrum for a class of oscillators}

\author{E. Sadurn\'i and E. Rivera-Moci\~nos}
\address{Instituto de F\'isica, Benem\'erita Universidad Aut\'onoma de Puebla,
Apartado Postal J-48, 72570 Puebla, M\'exico}
\ead{sadurni@ifuap.buap.mx, erivera@ifuap.buap.mx}

\begin{abstract}
We show that the position operator in a class of $f$-deformed oscillators has a fractal spectrum, homeomorphic to the Cantor set, via a unitary transformation to Harper's model. The class corresponds to a choice of ergodic operators for the deformation function. Hofstadter's butterfly is plotted by direct diagonalization of a position operator with an originally vanishing diagonal. This is equivalent to a one-dimensional hamiltonian without potential.
\end{abstract}





\section{Introduction}

The idea of deforming the oscillator algebra to obtain new solvable systems - in particular those connected to quantum optics \cite{manko1993, manko1995, manko1998, mancini2000} - is a vivid example of an old precept: there are two classes of problems, given by the unsolved ones and the harmonic oscillator; the idea is to move examples of the former class into the latter \cite{wolf2010}. In this respect, the $q$- and $f$- deformations of the Heisenberg algebra \cite{biedenharn1989, macfarlane1989} have proved to be sufficiently rich, containing hamiltonians with both finite and infinite levels in the spectrum \cite{atakishiev1990}. Moreover, the inherent nonlinearity of deformations suggests a strong relation to old, but relevant models of electrodynamics \cite{born1934, dirac1960, chung1965}. In a different realm of physics, but following the same precept, a mapping of a Dirac oscillator to a lattice has been reached by means of recursion relations \cite{sadurni2010}. This indicates a direct connection between the number of quanta of an oscillator and the site number of a semi-infinite chain \cite{franco2013}.  

Despite the vast literature related to this subject, little has been said about the configuration space emanating from algebraic deformations. In this paper we show the existence of an $f$-deformed Heisenberg algebra that turns the hamiltonian $H=\ahalf (P^2 + X^2)$ into an ergodic operator, while its position $X$ - equivalently $P$ - acquires a spectrum given by the Cantor set. This surprising result makes manifest the non-trivial topological properties of an algebraically modified space. For instance, the spectral measure of such a space vanishes, despite the uncountability of the eigenvalues associated to the deformed position operator. We thus have a {\it purely singular\ }phase space, if we borrow the term from spectral theory \cite{avron1990}.

We proceed in two steps. In section \ref{one} we show a direct relation between the position operator of a deformed Heisenberg algebra and a deformed tight-binding chain without on-site potentials. Then, in section \ref{two} we map unitarily an off-diagonal operator representing $X$ to Harper's model. The map contains as a special case a deformed semi-infinite tight-binding chain without potentials. The spectrum of the resulting system is found to be identical to Hofstadter's butterfly \cite{hofstadter1976} when plotted as a function of an external parameter. A few remarks on finite size effects in numerical results are given. We conclude in section \ref{conclusion}.

\section{A connection between the position of $f$-oscillators and a tight-binding chain without diagonal \label{one}}

The $f-$deformations of Heisenberg's algebra suggested by Man'ko \cite{manko1998} allow the introduction of a very general deformation function $F(N)$ of the usual number operator $N=\ahalf(p^2+x^2-1)$. In natural units, we have the well-known relations

\bea
\left[ x,p \right]=i, \qquad \left[ a, a^{\dagger} \right]=1, \qquad aF(N)=F(N+1)a.
\label{algebra}
\eea
\bea
a= \frac{1}{\sqrt{2}}(x+ip), \qquad a^{\dagger}= \frac{1}{\sqrt{2}}(x-ip).
\eea
Meanwhile, the deformed operators $A, A^{\dagger}$ are established\footnote{For convenience we use (\ref{deformed}), which is slightly different from the standard notation. See the third relation in (\ref{algebra}). } by means of $F(N)$ (hermitian) as

\bea
A= F(N) a, \qquad A^{ \dagger} = a^{\dagger} F^{\dagger}(N) =  a^{\dagger} F(N),
\label{deformed}
\eea
with the following modified commutations relations
\bea
\left[ A, A^{\dagger} \right] = F(N)^2 (N+1) - F(N-1)^2 N.
\eea
A deformed position $X$ and momentum $P$ can be constructed in the usual manner, but employing now $A, A^{ \dagger}$:

\bea
X= \frac{1}{\sqrt{2}}(A+A^{\dagger}), \qquad P = \frac{1}{i\sqrt{2}}(A-A^{\dagger}),
\eea
and the new commutation relation is
\bea
\left[X,P\right]= i \left[ F(N)^2 (N+1) - F(N-1)^2 N \right].
\eea
Evidently, these deformations have the effect of changing the spectrum of a hamiltonian conventionally defined as a quadratic form in phase space. We have

\bea
\fl H= \frac{1}{2} (P^2 + X^2) = \frac{1}{2} (AA^{\dagger} + A^{\dagger} A) = \frac{1}{2}  \left[ F(N)^2 (N+1) + F(N-1)^2 N \right],
\eea
showing that the spectrum $\sigma(H)$ is not necessarily equispaced. Since the last relation expresses $H$ as a function of $N$, $\sigma(H)$ can be obtained by acting directly on the usual harmonic oscillator basis $\{ | n \> \}_{n=0,1,2,...}$, resulting in the following expression

\bea
E_n = \frac{1}{2} \left[ F(n)^2 (n+1) + F(n-1)^2 n \right], \quad n=0,1,2,3,...
\eea
On the other hand, the spectrum of the position operator $X$ can be found by diagonalizing the matrix representation $\< n | X | m\>$ between harmonic oscillator states:

\bea
\< n | X | m\> = \frac{1}{\sqrt{2}} \left[ F(n) \sqrt{n+1} \delta_{n,m-1} + F(n-1) \sqrt{n} \delta_{n,m+1} \right].
\eea 
Incidentally, since phase space operators are unitarily related by

\bea
P = \exp \left(i \frac{\pi N}{2} \right) X \exp \left(- i \frac{\pi N}{2} \right) = i^N X i^{-N}, 
\eea
then $P$ and $X$ are isospectral. Now, if the eigenstates $| \xi \>$ of $X$ are expanded in the usual oscillator basis $| \xi \> = \sum_{n=0}^{\infty} \phi_n | n \> $, we are led to a recurrence relation for $\phi_n$ after substituting in $X |\xi\> = \xi | \xi \>$, i.e.

\bea
 \frac{F(n) \sqrt{n+1}}{\sqrt{2}} \phi_{n+1} + \frac{F(n-1) \sqrt{n}}{\sqrt{2}} \phi_{n-1} = \xi \phi_n
\eea
with the boundary condition $\phi_{-1} \equiv 0$. This can indeed be interpreted in terms of a nearest-neighbour tight-binding model: if we identify $F(n-1)\sqrt{n/2}$ with hopping amplitudes $\Delta_n$ (also known as couplings) and $\xi$ with the $k$-th energy of a hopping particle in a chain, we have the equivalent relation

\bea
 \Delta_{n+1} \phi^k_{n+1} + \Delta_n \phi^k_{n-1} = \varepsilon_k \phi^k_n
\eea
and the boundary condition can be fictitiously imposed by setting $\Delta_0 \equiv 0$; this completes the analogy with a semi-infinite chain. We may employ translation operators $T, T^{\dagger}$ and a site-number operator $N$ to express the tight-binding equation in hamiltonian form, 

\bea
 H_{t.b.}| \phi^k \>=\left[ \Delta(N) T+T^{\dagger} \Delta^{\dagger}(N) \right] | \phi^k \> = \varepsilon^k  | \phi^k \>, 
\eea
\bea
|\phi^k \> = \sum_{n=0}^{\infty} \phi_n^k | n \>,
\eea
with the properties
\bea
\fl T |n\> = | n+1 \>, \qquad T^{\dagger}|n\>= | n-1 \>,\qquad T^{\dagger} | 0 \>=0 ,\qquad N |n\> = n | n \>.
\eea
The atomic states $|n \>$ must now be interpreted as vectors whose space representations $\< y | n\>$ are functions of a real position variable $y$; such functions are localized around site $n=0,1,2,...$.
It is convenient to mention here that in certain applications of tight-binding models, such as photonic crystals \cite{russell2003} or optical lattices \cite{morsch2006, osterloh2005}, the coupling constants $\Delta$ can be engineered by means of evanescent waves coming out of a potential well or a resonating cavity located at a lattice site. 

\subsection{Ergodic operators for deformed oscillators}

Ergodic hamiltonians \cite{avila2012} are a key ingredient of fractal spectra, in particular for the `almost' Mathieu operator \cite{last1995}. In a more general fashion, an ergodic hamiltonian may contain operators whose spectrum is dense and uncountable in a subset of $\mathbb{R}$. Here we must specify our deformation function such that $\ahalf(P^2+X^2)$ is ergodic.

An operator $\hat O$ with such a property can be obtained by means of a quasiperiodic function of $N$. Let $\hat O$ be periodic of period $\tau$:

\bea
\hat O(N+\tau) = \hat O(N)
\eea
but now, if $\omega$ and $\tau$ are non-commensurate, the operator $\hat O(\omega N)$ does not have period $\tau$ in the variable $N$:

\bea
\hat O(\omega (N+\tau)) = \hat O(\omega N + \omega \tau) \neq \hat O(\omega N ).
\eea
Moreover, there is no translation of oscillator quanta that leads to invariance, i.e. for any $k, q \in \mathbb{Z}$

\bea
\hat O(\omega(N+k) ) = \hat O(\omega N + k \omega) \neq  \hat O(\omega N + q \tau) = \hat O(\omega N). 
\eea
In addition, if $\hat O$ is a hermitian bounded operator its spectrum must be dense, for acting on $|n\>$ yields a set equivalent to the irrational numbers within $\left[ O_{\scriptsize \mbox{min}} , O_{\scriptsize \mbox{max}}\right]$. A more specific example can be given in terms of trigonometric functions: If our deformation function $F$ is set as

\bea
F(N) = \frac{\sin \left[\omega (N+1) \right]}{\sqrt{N+1}} 
\label{f}
\eea
then the hamiltonian $H=\ahalf(P^2+X^2)$ will become

\bea
H = \frac{1}{2} \left(\sin^2 \left[\omega (N+1)\right] + \sin^2 \left[\omega N\right] \right).
\eea
Now we take $\omega$ to be non commensurate with $\pi$, such that $E_n$ generates a dense set contained in an interval. One can verify that the limits of the interval depend on $\omega$.

\section{A map between position operators and Harper's hamiltonian: the realization of Hofstadter's butterfly \label{two}}

The position operator $X$ now has the form of an ergodic operator, since it contains (\ref{f}) in its definition, and when $\omega$ is non commensurate with $\pi$, the spectrum of $\sin(\omega N)$ is dense in the interval $\left[-1,1\right]$. Here we claim and prove that in fact $X$ is isospectral to a tight-binding hamiltonian that produces Harper's equation, namely

\bea
H_{\scriptsize \mbox{Harper}} = T + T^{\dagger} + 2 \cos \left( \omega N - \nu \right).
\label{harper}
\eea
This is in principle unusual, since $H_{\scriptsize \mbox{Harper}}$ contains no lattice deformations (constant couplings) and the effect of fractality in its spectrum is due solely to the external field. On the other hand, $X$ is a pure deformation, without diagonal terms. We shall see that there exists a unitary $U$ such that $UXU^{\dagger}  = \lambda H_{\scriptsize \mbox{Harper}}$ with $\lambda$ an overall scaling factor.

Let us start with

\bea
X &=& \frac{\sin \left[ \omega(N+1) \right]}{\sqrt{2(N+1)}} a + a^{\dagger} \frac{\sin \left[ \omega(N+1) \right]}{\sqrt{2(N+1)}} \nonumber \\
&=&   a \frac{  \sin \left[ \omega N \right]}{\sqrt{2N}} +  \frac{\sin \left[ \omega N \right]}{\sqrt{2N}}  a^{\dagger}\nonumber \\
&=& a \frac{  \cos \left[ \omega N -\pi/2 \right]}{\sqrt{2N}} +  \frac{\cos \left[ \omega N -\pi/2 \right]}{\sqrt{2N}}  a^{\dagger}.
\eea
From here, we see that we may work with more generality using the following rescaled operator and the introduction of an extra parameter $\nu$

\bea
\fl X_{\nu} = a \frac{  2\cos \left[ \omega N -\nu \right]}{\sqrt{N}} +  \frac{ 2 \cos \left[ \omega N -\nu \right]}{\sqrt{N}}  a^{\dagger} \nonumber \\
\fl =  a \frac{  \exp \left[ i(\omega N -\nu) \right]}{\sqrt{N}} +  \frac{  \exp \left[- i(\omega N -\nu) \right]}{\sqrt{N}} a^{\dagger} +  a \frac{  \exp \left[- i(\omega N -\nu) \right]}{\sqrt{N}} +  \frac{  \exp \left[ i(\omega N -\nu) \right]}{\sqrt{N}} a^{\dagger} \nonumber \\
\fl \equiv H_0 + V,
\label{x}
\eea
where 
\bea
H_0= a \frac{  \exp \left[ i(\omega N -\nu) \right]}{\sqrt{N}} +  \frac{  \exp \left[- i(\omega N -\nu) \right]}{\sqrt{N}} a^{\dagger} \nonumber \\
V = a \frac{  \exp \left[- i(\omega N -\nu) \right]}{\sqrt{N}} +  \frac{  \exp \left[ i(\omega N -\nu) \right]}{\sqrt{N}} a^{\dagger}.
\eea
In what follows we show that $H_0$ corresponds to the addition of two translation operators and $V$ to a local potential of a transformed variable, using the appropriate basis.
In order to transform $X_{\nu}$ unitarily, we work now at the level of states using the definition

\bea
\fl | k ) = \frac{1}{\ncal} \sum_{n=0}^{\infty} \exp \left\{i \left[ \omega \left( \ahalf n (n+1) + k(k-1-2n) \right) + \nu n \right] \right\} | n \>, \qquad k \in \mathbb{Z},
\label{expansion}
\eea
with $\ncal$ an appropriate normalization constant. It is a matter of simple algebra to verify the following equations:

\bea
\fl \left\{\frac{  \exp \left[- i(\omega N -\nu) \right]}{\sqrt{N}} a^{\dagger} \right\}| k ) = | k+1 ), \quad 
 \left\{ a \frac{  \exp \left[ i(\omega N -\nu) \right]}{\sqrt{N}} \right\} | k ) = | k-1 )
\label{trans0}
\eea
\bea
\fl \left\{a \frac{  \exp \left[- i(\omega N -\nu) \right]}{\sqrt{N}} +  \frac{  \exp \left[ i(\omega N -\nu) \right]}{\sqrt{N}} a^{\dagger}\right\} |k) = 2 \cos\left[ 2 \omega k - 2\nu \right] | k ).
\label{trans}
\eea
Here one may proceed by direct substitution of (\ref{expansion}) in the second relation of (\ref{trans0}), while the first relation follows by recognizing that

\bea
|k+1) &=& \left\{\frac{  \exp \left[- i(\omega N -\nu) \right]}{\sqrt{N}} a^{\dagger} \right\}  \left\{ a \frac{  \exp \left[ i(\omega N -\nu) \right]}{\sqrt{N}} \right\} | k+1 ) \nonumber \\
&=&  \left\{\frac{  \exp \left[- i(\omega N -\nu) \right]}{\sqrt{N}} a^{\dagger} \right\} | k ).
\eea
In fact, the expansion coefficients in (\ref{expansion}) can be derived by solving the set of recurrences emerging from (\ref{trans0}, \ref{trans}) as requirements. Now that we have established these properties, the following identifications are in order

\bea
H_0 = T_k + T^{\dagger}_k, \qquad V = 2 \cos \left[ \tilde \omega K - \tilde \nu \right] 
\eea
i.e. $T_k, T^{\dagger}_k$ are translation operators in $k$ and $V$ is a function of the number operator $K$ such that $K |k)=k|k)$. We also write $\tilde \omega = 2 \omega$, $\tilde \nu = 2\nu$ to simplify the notation. Harper's equation now emerges transparently when $X_{\nu}$ is applied to its eigenvector $|\xi\>$ written as a linear superposition $|\xi \>=\sum_{k =-\infty}^{\infty} \psi_k |k)$; the recurrence is

\bea
\psi_{k+1} + \psi_{k-1} + 2\cos \left( \tilde \omega k - \tilde \nu\right) \psi_{k} = \xi \psi_{k}.
\eea
Moreover, since $k \in \mathbb{Z}$, this equation corresponds to an infinite chain.

It is left to verify that the transformation $|n\> \mapsto |k)$ is unitary. This is most easily done by applying the transpose conjugate of (\ref{expansion}) to $|k)$ as follows:

\bea
\fl \frac{1}{\ncal} \sum_{k=-\infty}^{\infty}  \exp \left\{-i \left[ \omega \left( \ahalf n (n+1) + k(k-1-2n) \right) + \nu n \right] \right\} |k) \nonumber \\
\fl = \frac{1}{\ncal}\sum_{k=-\infty}^{\infty}  \exp \left\{-i \left[ \omega \left( \ahalf n (n+1) + k(k-1-2n) \right) + \nu n \right] \right\} \nonumber \\ \times \frac{1}{\ncal}\sum_{m=0}^{\infty}  \exp \left\{i \left[ \omega \left( \ahalf m (m+1) + k(k-1-2m) \right) + \nu m \right] \right\} |m\> \nonumber \\
\fl =\sum_{m=0}^{\infty} \exp \left\{ i \left[ \ahalf \omega( m(m+1)-n(n+1)  ) + \nu (m-n)\right]  \right\} \left\{\frac{1}{\ncal^2}\sum_{k=-\infty}^{\infty} e^{2 i\omega k (n-m)}\right\}| m \> \nonumber \\
\fl = |n\>
\eea
where in the last line we recognize that the summation in braces is a properly normalized delta, so $m=n$.

It is interesting to note that the map described above does not work when (\ref{harper}) contains the more general potential $2\Lambda  \cos(\omega N- \nu)$ with $\Lambda \neq 1$. This seems to be connected with the known result \cite{avron1990} that only $\Lambda = 1$ produces a purely singular spectrum. For general $\Lambda$, the deformation function becomes complex

\bea
F(N) &=& \frac{\exp \left\{i \left[ \omega (N+1)-\nu \right] \right\}+ \Lambda \exp \left\{-i \left[ \omega (N+1)-\nu \right] \right\} }{\sqrt{N+1}} \nonumber \\
&=&  \frac{e^{i \Phi(N)}\sqrt{1+\Lambda^2 + 2\Lambda \cos\left[2 \omega(N+1)-2 \nu \right]}}{\sqrt{N+1}}
\eea
and its spectrum is modified. When $\Lambda \gg 1$ or $\Lambda \ll 1$, the trigonometric part becomes a mere perturbation.

\subsection{The spectrum of $X_{\nu}$}

\begin{figure}[t!]
\begin{center}  \includegraphics[width=16cm]{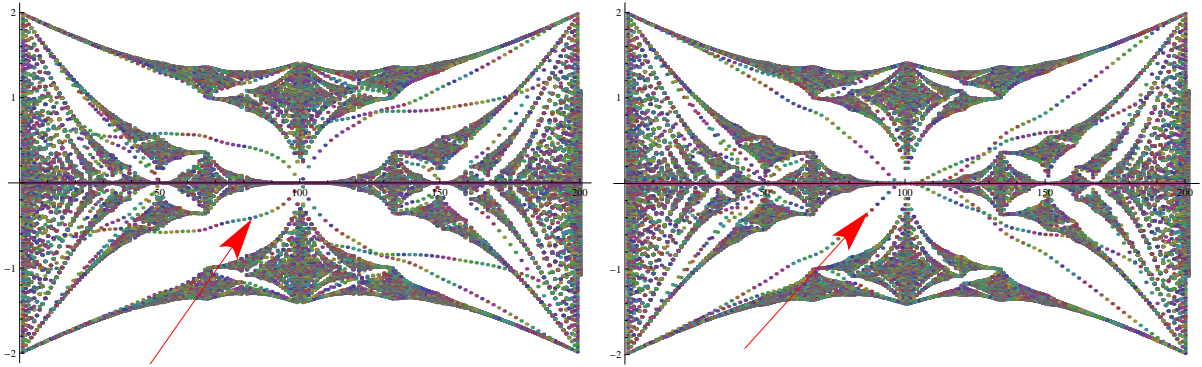} \end{center}
\caption{\label{fig:I.0} Numerical spectrum of two different off-diagonal ergodic operators $X_{\nu}$. Right panel: $n_{\scriptsize \mbox{max}}=200$. Left panel: $n_{\scriptsize \mbox{max}}=210$. The red arrows indicate a variation of edge states according to the size of the arrays. In both cases, the abscissa is partitioned in 200 fractions of $\pi$.}
\end{figure}

\begin{figure}[h!]
\begin{center}  \includegraphics[width=9cm]{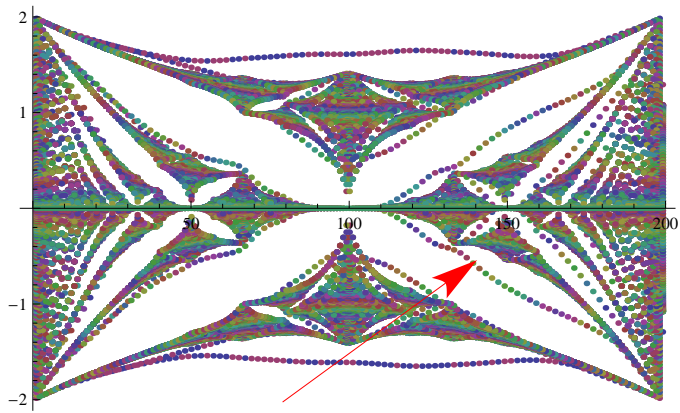} \end{center}
\caption{\label{fig:I.1} Numerical spectrum of the ergodic operator $X_{\nu}$, $n_{\scriptsize \mbox{max}}=200$ and periodic boundary conditions. The red arrow shows the location of edge states, at their new position.}
\end{figure}

As pointed out in very early works \cite{hofstadter1976, azbel1964} on this subject, the spectrum of this system is equivalent to Cantor's set when $\omega$ is non commensurate with $\pi$, i.e. it is purely singular. This is independent of the value of $\nu$. On the other hand, if $\omega$ and $\pi$ are commensurate we have $q\omega = p\pi$, and the system becomes periodic. The period is $q$ if $p,q$ are relatively primes; this implies that a structure of $q$ bands should emerge. As $p/q$ approaches irrationality, the number of such bands increases as a power of a given initial number (say, two bands for a precursor of Cantor's set) and their length decreases in a similar amount. This analysis is also valid for our off-diagonal $X_{\nu}$ as expressed in (\ref{x}), since it only requires the ergodic behavior of $\sin(\omega N + \nu)$. Usually, the spectrum is depicted by showing a sequence of shrinking intervals as a function of $\omega$, giving rise to the famous butterfly. In our problem, we may diagonalize directly the operator $X_{\nu}$ by a suitable numerical routine and plot the spectrum. Since the matrix representations that we use are of a finite size, an additional artefact appears in the pattern: The gaps between bands for rational $\omega$ contain eigenvalues due to edge states. Such bands proliferate in a self-similar pattern and it is to be expected that edge states appear in various regions of the spectral span. To verify that indeed such states are due to finite size effects, we proceed to diagonalize $X_{\nu}$ for increasing sizes $n_{\scriptsize \mbox{max}}=200, 210, ...$ and we vary $\omega=\pi/200, 2\pi/200, 3\pi/200,...$ for each size. We show the results in fig. \ref{fig:I.0}. In fig. \ref{fig:I.1} we repeat the calculation, but with periodic boundary conditions; the results reveal a significant displacement of border levels towards the allowed bands.

\section{Conclusion \label{conclusion}}

A fractal position spectrum has been found. The example that we have studied has been shown to be equivalent to the reduced problem of an electron in a square lattice under the influence of a magnetic field. We also computed the spectrum numerically to verify the property of isospectrality, recognizing that very small discrepancies arise due to finite size. Undoubtedly, experiments can be carried out in the context of artificial solids, such as those produced by optical lattices, photonic crystals or microwave resonators. In particular, a nearest-neighbour tight-binding chain of identical sites and adjustable distances has been used succesfully to produce other models, and this choice is likely to give immediate results in the present case. It is also important to note that a realization of ergodic hamiltonians using a fixed photon mode in a cavity is not a complete diversion from quantum optical applications: it is a non-linear example of oscillation that can be studied and perhaps partially tested. Due to the extreme dependence of its effective frequency $\Omega(n) = E_n/( n+1/2)$, we may propose $\Omega$ to vary randomly with the occupation number. 

\ack

We are grateful to CONACyT for financial support under project CB2012-180585. E.R.-M. also wishes to thank CONACyT for {\it beca-cr\'edito\ }245104.

\section*{References}

\end{document}